# Positron annihilation lifetime and Doppler broadening spectral calculations of oxygen-doped 3*C*-SiC

Yi Zhao[1], Hongtao Zhang[1], Qiang Li[2], Xian Tang[1,*], Guodong Cheng[2,*]

1.School of Nuclear Science and Technology, University of South China, Hengyang 421001, China

2.School of Computer, University of South China, Hengyang 421001, China

*Emails: xiantang@usc.edu.cn (X.T.); chenggd@usc.edu.cn (G.C.)

**Abstract**

Based on density functional theory (DFT), the formation energies of intrinsic vacancy defects ($V_C$, $V_{Si}$, and $V_{Si+C}$) and oxygen-related defects ($O_C$, $O_{Si}$, $O_C V_{Si}$, and $O_{Si} V_C$) in 3*C*-SiC are systematically investigated. The results indicate that all defects considered, except for $O_C$, possess neutral or negative charge states, thereby making them suitable for detection by positron annihilation spectroscopy (PAS). Furthermore, the electron and positron density distributions and positron annihilation lifetimes for the perfect 3*C*-SiC supercell and various defective configurations are computed. It is found that the $O_{Si}$ and $O_{Si} V_C$ complexes act as effective positron trapping centers, leading to the formation of positron trapped states and a notable increase in annihilation lifetimes at the corresponding defect sites. In addition, coincidence Doppler broadening (CDB) spectra, along with the *S* and *W* parameters, are calculated for both intrinsic and oxygen-doped point defects ($O_C$, $O_{Si}$, $O_C V_{Si}$, and $O_{Si} V_C$). The analysis reveals that electron screening effects dominate the annihilation characteristics of the $O_{Si}$ defect, whereas positron localization induced by the vacancy is the predominant contributor in the case of $O_{Si} V_C$. This distinction results in clearly different momentum distributions of these two oxygen-related defects for different charge states. Overall, the PAS is demonstrated to be a powerful technique for distinguishing intrinsic vacancy-type defects and oxygen-doped composites in 3*C*-SiC. Combining the analysis of electron and positron density distributions, the electron localization and positron trapping behavior in defect systems with different charge states can be comprehensively understood. These first-principles results provide a solid theoretical foundation for identifying and characterizing the defects in oxygen-doped 3*C*-SiC by using PAS.

# 1. Introduction

3*C*-SiC is a kind of cubic silicon carbide, which has low density, high strength, excellent thermal stability, high thermal conductivity and excellent neutron irradiation resistance. It has been widely used in the fields of fuel particle cladding of high temperature gas-cooled reactor, substrate materials of gas-cooled fast reactor and structural materials of molten salt reactor[1–3]. However, the introduction of oxygen impurities is unavoidable in the preparation of 3$C$-SiC, and its source is closely related to the process conditions and material properties. First, silicon sources (such as $SiH_4$ and $SiCl_4$) and carbon sources (such as $CH_4$ and $C_3H_8$) often contain trace oxides or adsorbed oxygen, which are easy to release oxygen atoms through decomposition or oxidation side reactions in high temperature reactions (>1200°C)[4].

Secondly, in the chemical vapor deposition (CVD) or physical vapor transport (PVT) process, the quartz reaction chamber ($SiO_2$) is corroded in the reducing gas ($H_2$), resulting in the incorporation of gaseous oxygen into the lattice. In addition, the residual $H_2O/O_2$ in the ambient humidity or protective gas (Ar and $N_2$) decomposes at high temperature, which further aggravates the oxygen pollution[5]. Oxygen atoms tend to occupy the interstitial or substitutional sites of SiC lattice, forming Si-O-C complex defects and $SiO_2$ nano-precipitates, which significantly reduce the carrier mobility and induce the increase of interface state density. However, oxygen impurities can lead to the decrease of high temperature resistance, corrosion resistance, creep resistance and neutron irradiation resistance of SiC[6–8], and eventually lead to material failure. At the same time, the high temperature, high pressure and high radiation environment of the reactor will produce a large number of high-energy particles to impact the material lattice, which will cause atomic displacement and point defects. Therefore, it is of great significance to explore the effect of oxygen doping on the microstructure defects of 3$C$-SiC materials for its application in the field of nuclear materials.

In recent years, Rosso and Baierle[9] have studied the stability and electronic properties of oxygen doped SiC nanowires by first-principles calculations. It is found that oxygen is more stable at the carbon site ($O_C$) than at the silicon site ($O_{Si}$), and oxygen impurities tend to migrate to the nanowire surface. Electronic structure calculations show that oxygen doping introduces electronic states into the band gap, especially when oxygen saturates the surface dangling bonds, which shows spin-dependent electronic properties and leads to the generation of magnetic moments. These results reveal a significant effect of oxygen on the electronic structure of SiC nanowires. At the same time, there are few studies on the identification of O-doped 3$C$-SiC. Gali et al.[10] studied the point defects of oxygen in 3$C$-SiC and 4$H$-SiC by first-principles calculations. The results show that the $O_C$ behave as double donors, and the $O_C$ defect in 3$C$-SiC is more stable than $O_{Si}$ or $O_i$ (interstitial oxygen).

Positron Annihilation Spectroscopy (PAS)[11,12] is a non-destructive technique for detecting the internal microstructure of materials based on the process by which positrons and electrons interact and annihilate to produce gamma rays ($\gamma$-photons). The annihilation lifetime of a positron can be obtained by accurately measuring the time interval between the incident positron and the gamma photon produced by its annihilation with an electron. In an ideal lattice structure, because of the uniform electron density between atoms, positrons enter the material in a "free state annihilation". However, when there are defects such as vacancies and dopant atoms in the material, these defects will cause local changes in the electron density, which may lead to the capture of positrons by vacancies and prolong the annihilation lifetime compared with the ideal lattice. PAS has been widely used to study vacancies and doped defects[13–19].

In recent years, Lam et al.[20] have studied the defect behavior of n-type 6*H*-SiC under 8 MeV electron irradiation, discovered the existence of carbon-silicon divacancy by positron lifetime technique, and discussed the change of positron trapping after annealing at 1200 ℃. At the same time, Staab et al. Have studied various defects in 4*H*-SiC and 6*H*-SiC[21], including carbon vacancy, silicon vacancy and carbon-silicon divacancy. By calculating the positron annihilation lifetime and coincidence Doppler broadening (CDB)[22] of these defects, the effects of different defects on the positron trapping behavior are revealed, especially the positron lifetime extension effect. The microscopic properties of these defects are particularly pronounced under irradiation damage conditions. Brauer et al.[23] used Doppler broadening and PAS to study the radiation damage of 6*H*-SiC implanted with 200 keV Ge$^+$. The results showed that divacancies and silicon monovacancies were the main defects, and vacancy clusters were formed in the amorphous layer. In addition, Brauer et al.[24] further evaluated the basic defect characteristics in 6*H*-SiC by PAS, and found that the vacancy clusters in the amorphous layer after irradiation were more significant, which revealed the law of defect evolution induced by irradiation. Kawasuso et al.[25] studied the thermal evolution of carbon and silicon vacancies in 4*H*-SiC after electron irradiation, and discussed the elimination mechanism of defects at different temperatures. Hu et al.[26] analyzed the relationship between the formation of vacancy clusters and the expansion of 3*C*-SiC after neutron irradiation by PAS, and revealed the influence of these defects on the structural stability of 3*C*-SiC. In addition, Hu et al.[26] used CDB measurements to study the chemical characteristics around the positron capture site. Defects associated with silicon vacancies were found to be dominant in the materials studied.

Based on first-principles calculations [27–29], this study investigates 3*C*-SiC with the space group *F*−43*m* using the meta-GGA HLE17 functional scheme [32] within the generalized gradient approximation (GGA) [30,31]. In this structure, Si atoms occupy the Wyckoff 4a sites and C atoms occupy the 4c sites. First, the formation energies of various charged defects in 3*C*-SiC were calculated, and the stability of O-doped point defects as a function of charge state and Fermi level was examined. Second, the electron–positron densities were computed for the neutral bulk state as well as for neutral and negatively charged $O_{Si}$ and $O_{Si}V_C$ defects. Third, four different positron annihilation lifetime calculation schemes were employed to determine the positron annihilation lifetimes for each neutral and negatively charged defect, and the results were compared with experimental lifetimes reported in the literature. Finally, the CDB spectra and the *S* and *W* parameters were calculated for intrinsic point defects and O-doped point defects ($O_C$, $O_{Si}$, $O_C V_{Si}$ and $O_{Si} V_C$), providing further analysis of their electron–positron momentum distributions.

## 2. Calculation method

The calculation is based on density function theory (DFT)[33,34], and the projector augmented wave (PAW) [35] density function theory method is used to describe the interaction between the ionic core and the valence electron; The exchange-correlation interaction is calculated by using the parameters of the HLE17 scheme of the meta-GGA functional in the generalized gradient approximation, which is used to give consideration to both accuracy and efficiency. It is verified by Rauch et al.[36] that the accuracy of the meta-GGA functional in calculating defect energy levels and band gaps is close to that of the HSE06 hybrid functional, and the computational cost is much lower than that of the HSE06 hybrid functional; The defect formation energy, positron annihilation lifetime, differential charge density and CDB spectrum of 3*C*-SiC are calculated by expanding the electronic wave function with a plane-wave basis set. The plane wave cutoff energy is set to 450 eV, and the Brillouin zone is divided into $3 \times 3 \times 3$ *k* point networks for integration using the Gamma method. In the structural optimization, the convergence criterion of residual stress was set to be 0.01 eV/Å. In the electronic self-consistent iteration, the convergence condition is $1 \times 10^{-6}$ eV. A $3 \times 3 \times 3$ 3*C*-SiC supercell containing 216 atoms was constructed for defect calculation. The spin polarization effect is considered in the whole calculation, and the lattice constant and atomic position of the material are fully optimized. The calculated lattice parameter is 4.16 Å, which is close to the experimental value of 4.36 Å[37]. The calculated band gap is 2. 13 eV, which is close to the experimental value of 2. 36 eV[38]. The defect formation energies and electronic structures involved in this paper are calculated by VASP software.

The formation energy of the valence defect of the material can be calculated by the following equation:

$$E_f(V_X, q) = E_{tot}(V_X, q) - E_{SiC}^{bulk} - \sum_i n_i \mu_i + q(E_{VBM} + \mu_e + \Delta V), \quad (1)$$

Where $E_{SiC}^{bulk}$ denotes the total energy after perfect supercell relaxation; $V_X$ represents vacancy defect of *X* element (*X* is Si or C); $E_{tot}(V_X, q)$ represents the total energy after relaxation of the corresponding defect in the valence state of *q*; $\sum_i n_i \mu_i$ is a summation term representing the energy change when atoms are removed or added from the system, $n_i$ is the number of atoms of the $i$ species (negative for removal and positive for addition), and $\mu_i$ is the energy of the $i$ species after relaxation; $E_{\text{VBM}}$ represents the energy of the top of the valence band; $\mu_e$ represents the value of the Fermi level; $\Delta V$ is a correction term used to correct the electrostatic potential energy in the lattice with defects. By introducing the correction term $\Delta V$, the deviation of the electrostatic potential energy caused by the system defects can be corrected.

The positron annihilation lifetime is calculated by two-component density functional theory [39,40], in which the polarization effect is taken into account and the interaction between electrons and positrons is described by an electron-positron correlation potential function. There is no need to directly solve the complex many-body problem associated with positron-electron annihilation. The calculation of positron annihilation lifetime is divided into two steps: first, the ground state electron density is calculated with high accuracy. Then, fixing the electron density distribution, the positron Kohn-Sham equation is solved to obtain the positron density. The positron annihilation lifetime is the inverse of the positron annihilation rate, which is calculated by

$$\frac{1}{\tau} = \lambda = \pi c r_0^2 \int_{R^3} d^3\mathbf{r}\, n^+(\mathbf{r}) n(\mathbf{r}) g(n^+, n), \tag{2}$$

where $\lambda$ denotes the positron annihilation rate, and $\lambda$ is the inverse of the positron lifetime $\tau$; The $c$ represents the speed of light; $r_0$ denotes the classical radius of the electron; $n^+(\mathbf{r})$ and $n(\mathbf{r})$ represent positron density and electron density, respectively; $g(n^+, n)$ is the enhancement factor. The core challenge of positron annihilation lifetime calculation is to accurately describe the interaction between positrons and electrons. Generally, the description methods of this interaction are divided into two main categories: local density approximation (LDA)[34] and GGA. In the LDA scheme, the BNLDA proposed by Boronski and Nieminen[40] is based on the numerical results of the multi-particle random phase approximation (RPA)[41] and the numerical results of Nieminen et al.[39]. This model provides an accurate description of the electron-positron correlation energy and enhancement factor in the low electron density region. In the GGA scheme, the APGGA scheme proposed by Barbiellini et al.[42] improves the description of the electron-positron correlation energy by introducing the effect of the electron density gradient. The QMCGGA[43] scheme, based on quantum Monte Carlo (QMC) simulations, corrects GGA by using the results of QMC simulations. The perturbed hypernetted chain (PHNC) formula based PHNCGGA scheme proposed by Boronski[44] provides a new method for calculating the electron-positron correlation energy by integrating the PHNC formula into GGA.

Positron Doppler broadening spectroscopy can be used to identify elements in positron annihilation (first proposed by Asoka-Kumar et al.[45]), to detect changes in lattice structure (first proposed by Alatalo et al.[46]), etc. In fact, Doppler broadening technology has been used to detect defects in materials[47–50]. The theoretical calculation of positron Doppler broadening spectrum can be divided into two frameworks: self-consistent field calculation and non-self-consistent field calculation.

For the calculation of positron Doppler broadening, the state-independent momentum density distribution model is used, and the positron-electron momentum density distribution is[51–53]

$$\rho(\boldsymbol{p}) = \sum_{i\boldsymbol{k}} f(\varepsilon_{i\boldsymbol{k}}) \left| \int \mathrm{d}\boldsymbol{r} \exp(-\mathrm{i}\boldsymbol{p}\cdot\boldsymbol{r})\psi_{+}(\boldsymbol{r}) \times \psi_{i\boldsymbol{k}} g(n^{+}, n) \right|^2, \tag{3}$$

Where $\psi_{+}(\boldsymbol{r})$ is the positron ground state wave function, $\varepsilon_{i\boldsymbol{k}}$ is the $i$ band, the wave vector is the electronic state energy of $\boldsymbol{k}$, $\psi_{i\boldsymbol{k}}(\boldsymbol{r})$ is the electron wave function, and $f$ is the electron occupation number. In Eq. 3 the sum over all occupied electronic states is the electron-positron three-dimensional momentum distribution density. The Doppler broadening spectrum can be obtained by integrating the momentum density distribution in two dimensions:

$$N(p_z) = \mathrm{Const} \times \int \rho(\boldsymbol{p}) \mathrm{d}p_x \mathrm{d}p_y. \tag{4}$$

**3. Results and Discussion**

**3.1 Defect formation energy**

The thermodynamic stability of defects in materials can be effectively evaluated by the calculation of defect formation energy, and the formation tendency of different defects and their corresponding charge states can be directly reflected. Considering the various possible charge states of defects, the formation energies of defects in 3*C*-SiC with charge states ranging from − 2 to + 2 were calculated under the carbon-rich limit.

Fig. 1 shows the calculated defect formation energy of 3*C*-SiC as a function of the electron chemical potential. The electron chemical potential ranges from 0 to 2.13 eV. The calculated results of intrinsic defects are in good agreement with those of Wiktor et al.[54]. From Fig. 1, we know that: 1) Among the intrinsic defects, the $V_C$ is stable only in the + 2 and 0 valence states, and only close to the conduction band minimum at the Fermi level. However, $V_{Si}$ and $V_{Si+C}$ are stable in + 1, − 1, − 2 and + 1,0, − 2 valence states, respectively, and their formation energy curves are always higher than $V_C$, indicating that $V_C$ is the most thermodynamically stable for intrinsic defects; 2) Among the oxygen-related defects, $O_{Si}$ is stable in all valence states and has a low formation energy. $O_CV_{Si}$ is stable in the +1, 0, −1 and −2 valence states, and $O_{Si}V_C$ is stable in the considered valence states. In addition, $O_C$ does not have a stable neutral or negative valence state, which is not shown in Fig. 1. The analysis of Fig. 1 shows that there are stable neutral or negative valence states in oxygen-related complex defects except $O_C$, and there are stable neutral or negative valence states in Si-related defects ($V_{Si}$, $V_{Si+C}$), which is consistent with the conclusion of Hu et al.[26]. The calculation results show that all the defects considered except $O_C$ have neutral or negative valence states, which are suitable for PAS detection.

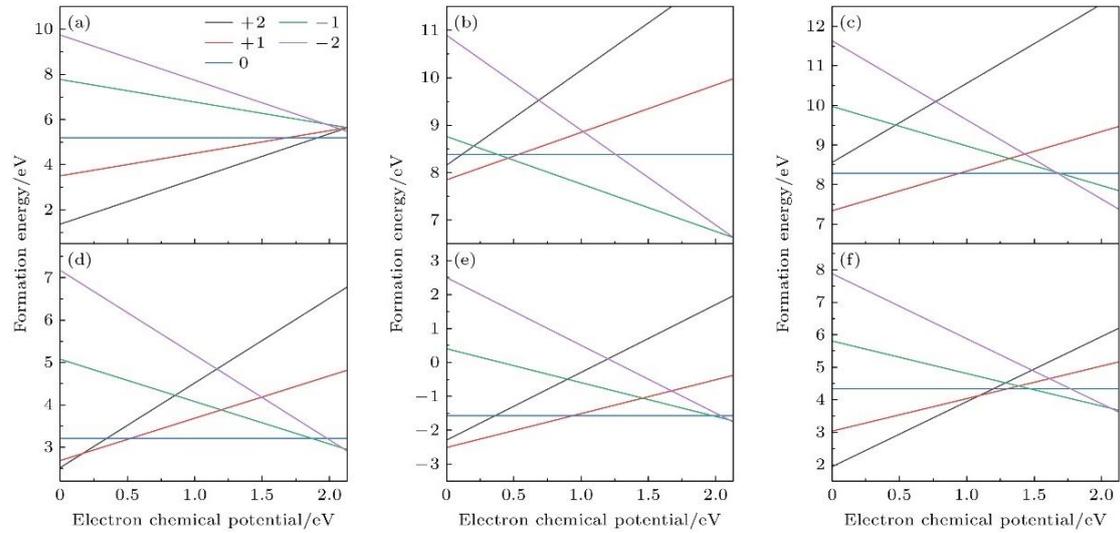

**Figure 1.** Formation energies of intrinsic and oxygen-doped defects in 3$C$-SiC: (a) $V_C$; (b) $V_{Si}$; (c) $V_{Si+C}$; (d) $O_{Si}$; (e) $O_CV_{Si}$; (f) $O_{Si}V_C$.

**3.2 Electron-positron density distribution.**

Before the calculation of positron lifetime, the electron-positron density distributions of the neutral state and the negative valence state of the defect-free crystal and the $O_{Si}$ and $O_{Si}V_C$ in 3$C$-SiC were studied. As shown in Figs. 2 and 3, the horizontal and vertical axes represent the number of grid points in real space. The red line represents the positron density, while the background color gradient from light to dark represents the electron density change from low to high.

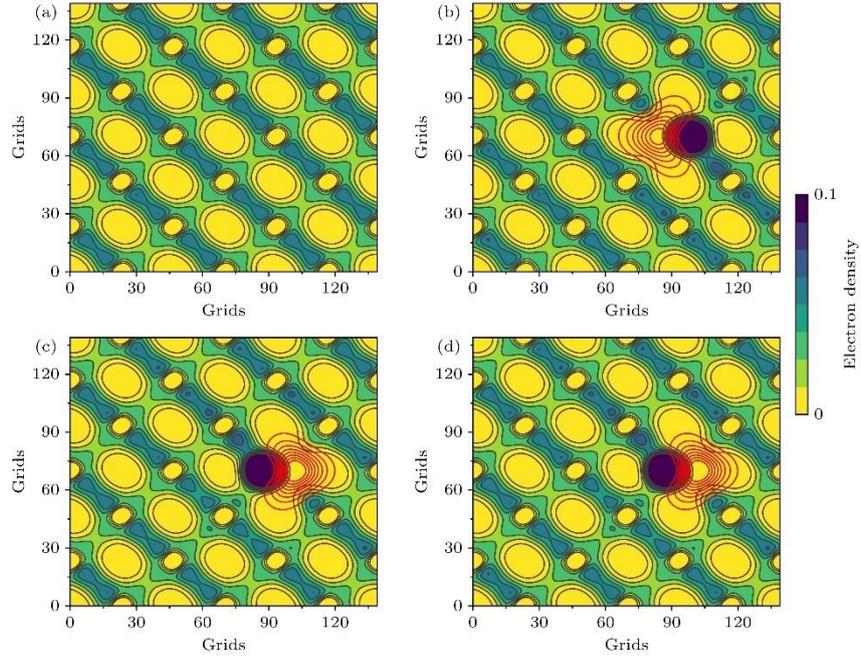

**Figure 2.** Electron-positron density: (a) Defect-free crystal; (b) neutral state of $O_{Si}$; (c) −1 charge state of $O_{Si}$; (d) −2 charge state of $O_{Si}$

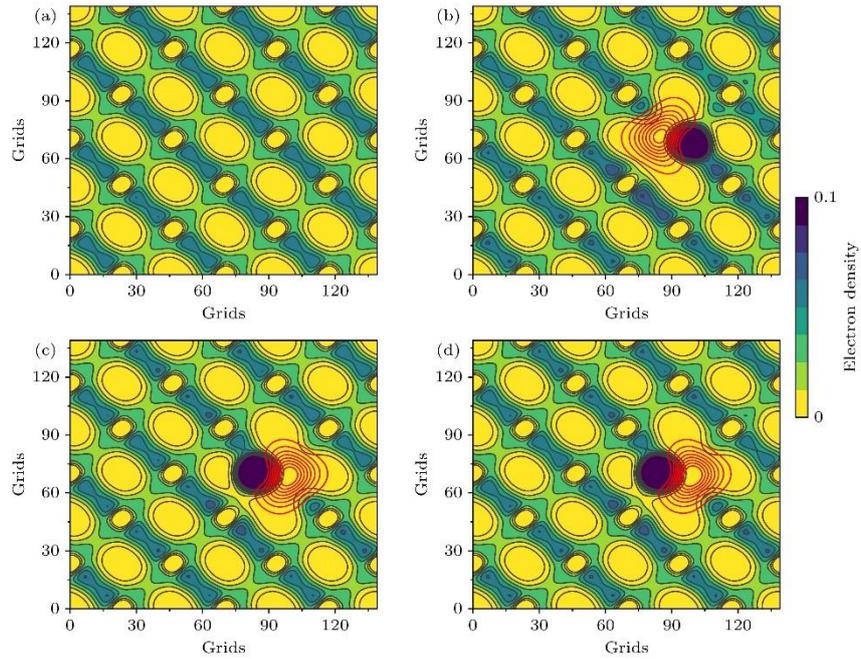

**Figure 3.** Electron-positron density: (a) Defect-free crystal; (b) neutral state of $O_{Si}V_C$; (c) −1 charge state of $O_{Si}V_C$; (d) −2 charge state of $O_{Si}V_C$.

In the defect-free bulk system, the electron density presents a regular periodic distribution, which is mainly localized around the C atom, and the positron is mainly distributed in the interstitial of SiC, and does not show localization, indicating that the positron is mainly annihilated in the free state in the ideal crystal structure. When $O_{Si}$ and $O_{Si}V_C$ defects are introduced into the system, a significant positron density distribution appears at the neutral state defect site, and the electron density distribution is also locally distorted due to the existence of defects. Combined with the positron lifetime data, the significant difference between the neutral state and the negative valence state (– 1 and – 2) is essentially due to the difference in the distribution of the defect charge state to the surrounding electron density in different valence states. The charge difference between the neutral state and the negative valence state will change the distribution[55] of the electron density and the positron density around the defect, resulting in the difference in the overlap integral between the electron density and the positron density, thereby changing the positron annihilation lifetime of the system. Further analysis of $O_{Si}$ and $O_{Si}V_C$ defect negative valence states (– 1 and – 2) shows that there is almost no difference in positron lifetime between them. From Fig. 2(c),(d) and Fig. 3(c),(d), it can be seen that the positron density distribution and electron density distribution in the two negative valence states are less affected by the defect, so the lifetime is close to. In addition, the positron lifetimes in the three valence states are larger than those in the defect-free bulk system, indicating that the $O_{Si}$ and $O_{Si}V_C$ defects act as positron capture centers to form positron capture state annihilation, which prolongs the positron annihilation lifetime at the defects.

### 3.3 Positron annihilation lifetime

The analysis of defect formation energy by Fig. 1 shows that there are stable neutral and negative valence defects in the studied defects, so the positron annihilation lifetimes of these defects and perfect supercells are further calculated. The positron annihilation lifetime and the local magnetic moment of the defect calculated according to the four electron-positron correlation models are listed in the Table 1.

Table 1. Calculated positron annihilation lifetimes (ps) for the four schemes.

| Type | BNLDA | APGGA | PHNCGGA | QMCGGA | Document |
|---|---|---|---|---|---|
| Bulk phase | 150 | 150 | 147 | 153 | 145[56] |
| $V_C^0$ | 151 | 150 | 147 | 152 | 150[23] |
| $V_{Si}^0$ | 241 | 238 | 233 | 242 | 227[54] |
| $V_{Si}^{1-}$ | 237 | 233 | 229 | 238 | 225[54] |
| $V_{Si}^{2-}$ | 236 | 232 | 228 | 237 | 222[54] |
| $V_{Si+C}^0$ | 250 | 249 | 243 | 251 | |
| $V_{Si+C}^{1-}$ | 243 | 242 | 236 | 245 | |
| $V_{Si+C}^{2-}$ | 239 | 244 | 235 | 242 | |
| $O_{Si}^0$ | 164 | 170 | 164 | 169 | |
| $O_{Si}^{1-}$ | 167 | 187 | 175 | 176 | |
| $O_{Si}^{2-}$ | 167 | 187 | 174 | 175 | |
| $O_C V_{Si}^0$ | 239 | 242 | 234 | 242 | |
| $O_C V_{Si}^{1-}$ | 237 | 242 | 234 | 240 | |
| $O_C V_{Si}^{2-}$ | 234 | 240 | 231 | 238 | |
| $O_{Si} V_C^0$ | 181 | 186 | 180 | 186 | |
| $O_{Si} V_C^{1-}$ | 183 | 202 | 190 | 192 | |
| $O_{Si} V_C^{2-}$ | 183 | 202 | 190 | 191 | |

From Table 1, it can be seen that the positron annihilation lifetime is 147-153 ps in the defect-free bulk structure, and the calculated positron lifetime is close to the experimental value of 145 ps measured by Panda et al.[56]; The positron lifetime of the V0C is 147-152 ps, which is consistent with the experimental value of 150 ps measured by Brauer et al.[23]; The positron lifetime of $V_{Si+C}$ is quite different from the reference value. By comparing the calculated positron lifetime, the lifetime of $V_C$ is close to that of the bulk phase, indicating that the local electron density changes little; The lifetimes of $V_{Si}$ and $V_{Si+C}$ are significantly prolonged and increase with the increase of vacancy complexity, which reflects the significant influence of intrinsic defects on the annihilation behavior. In contrast, the lifetime of $O_C$ is almost the same as that of the bulk due to the limited change of the local electron density caused by the acceptor characteristics, while the lifetime of $O_{Si}$ is higher than that of the bulk, reflecting the decrease of the electron density caused by the donor repulsion and the deep level trap, and the lifetime of $V_{Si}$ is lower than that of $V_{Si}$. Among the oxygen recombination defects, the lifetime of $O_C V_{Si}$ is close to but slightly lower than that of $V_{Si+C}$, which needs to be further distinguished by its negative charge characteristics, while the lifetime of $O_{Si} V_C$ is greater than that of $O_{Si}$ and $V_C$, highlighting the complex modulation of the electronic structure by the synergistic effect of oxygen doping and vacancies. Oxygen-doped defects can be clearly identified by lifetime difference and charge state analysis. The positron lifetime calculated in Table 1 is quite different from the experimental reference value on the whole, which may be caused by the consideration of spin polarization effect in the calculation process and the additional consideration of orbital kinetic energy density in meta-GGA functional. The positron annihilation process follows the spin selection rule, that is, the annihilation mainly occurs in the singlet pair composed of the opposite spin of the positron and the electron. When there is spin polarization (such as magnetic defects or ferromagnetic regions) in the material, the spin polarization of electrons will change the charge density distribution of the local system, thus affecting the positron annihilation lifetime[57]. Compared with the traditional GGA functional, the meta-GGA functional not only depends on the electron density and its gradient, but also senses the local variation of the electron wave function by introducing a kinetic energy density term into the exchange-correlation energy density, in which the kinetic energy density reflects the heterogeneity and localization of electrons in space, thus better distinguishing core electrons from valence electrons, and more accurately describing the local behavior of electrons, thus affecting the calculation of positron annihilation lifetime (positron lifetime). On $O_{Si}$ and $O_{Si}V_C$, it can be observed that the positron lifetime difference between the neutral state and the negative valence state is large, but the positron lifetime difference between the negative valence states (– 1 and – 2) is small, which is consistent with the calculation results of the electron-positron density distribution. The positron lifetime in the neutral state of $O_{Si}$ is less than that in the negative valence state, while the positron lifetime in the neutral state of $O_C V_{Si}^0$ is greater than that in the negative valence state, which is consistent with the calculation results of the subsequent Doppler broadening spectrum. In addition, the magnetic moments of $V_C^0$ and $O_{Si}^0$ are – 2 A ·m$^2$, and the magnetic moments of $V_{Si}^0$, $V_{Si+C}^0$ and $O_C V_{Si}^0$ are 2 A ·m$^2$, indicating that there are two unpaired electrons and significant spin polarization.

In this study, the effects of different defect types on the electronic structure of materials were systematically analyzed by positron lifetime calculation. As listed in Table 1, the calculated bulk positron lifetime is slightly larger than the experimental value of Panda et al.[56], while the calculated monovacancy $V_C^0$ is highly consistent with the experimental reference value of Brauer et al.[23,24]. It is worth noting that the calculated results of $V_{Si+C}$ composite vacancy are significantly different from the experimental values, which may be due to the consideration of spin polarization effect and the additional consideration of orbital kinetic energy density by meta-GGA functional.

The defect type analysis shows that the positron lifetime of the monovacancy $V_C$ is similar to the bulk lifetime, indicating that the local electron density distribution of the monovacancy $V_C$ is not significantly changed. In contrast, the lifetimes of monovacancy $V_{Si}$ and its complex vacancy $V_{Si+C}$ show an obvious increasing trend, indicating that the increase of intrinsic defect complexity will delay the positron annihilation behavior. Among the oxygen doped defects, the $O_C$ defect is limited by the change of local electron density, and its lifetime is basically the same as that of the bulk phase; On the other hand, the electron density of $O_{Si}$ donor defect is reduced due to the dual effects of charge repulsion and deep level traps, but its lifetime is still lower than that of monovacancy V due to the stronger lattice distortion effect of $V_{SiSi}$.

The study of oxygen recombination defect system shows that the lifetime of $O_CV_{Si}$ recombination defect is close to but slightly lower than that of $V_{Si+C}$, which may be related to the difference of electron density distribution caused by its negative charge characteristic. It is worth noting that the lifetime of the $O_{Si}V_C$ complex defect is significantly higher than that of the $O_{Si}$ and $V_C$ single defect system, which is due to the synergistic effect of oxygen doping and vacancy defects. Charge state analysis further reveals that the positron lifetime of the $O_{Si}$ defect in the negative charge state is higher than that in the neutral state, while the neutral state lifetime of the $O_CV_{Si}$ defect is higher than that in the negative charge state, which is consistent with the Doppler broadening spectrum calculation.

**3.4 Coincidence Doppler broadening spectrum**

The results of CDB can reveal how these defects affect the electronic structure of materials, especially in the electronic energy levels. All Doppler momentum spectra presented in this section are set with an integration window of $S \in (0, 2.86) \times 10^{-3} m_0c$ and $W \in (10.58, 27.36) \times 10^{-3} m_0c$, convolved using a Gaussian function with full-width-at-half-maximum (FWHM) = $4.7 \times 10^{-3} m_0c$, and the results shown are $(0 - 40) \times 10^{-3} m$ UN. The calculated Doppler momentum spectra are shown in Figs. 4 and 5.

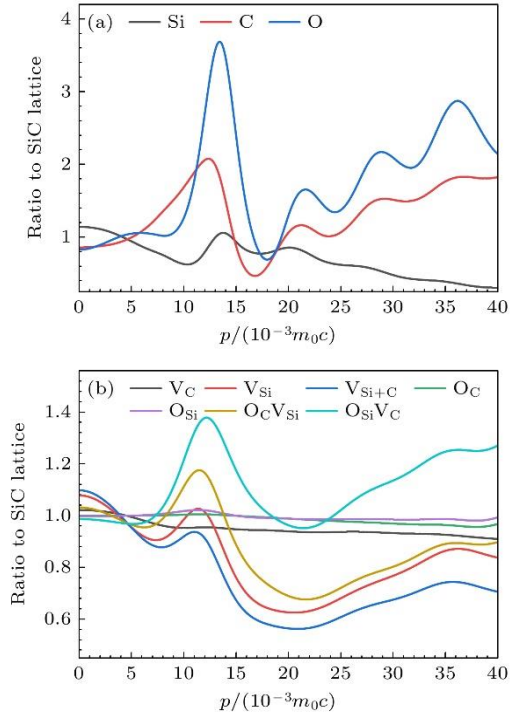

**Figure 4.** (a) Doppler spectra of the C, Si, and O; (b) Doppler spectra of the $V_C$, $V_{Si}$, $V_{Si+C}$, $O_C$, $O_{Si}$, $O_CV_{Si}$ and $O_{Si}V_C$ defects in 3C-SiC.

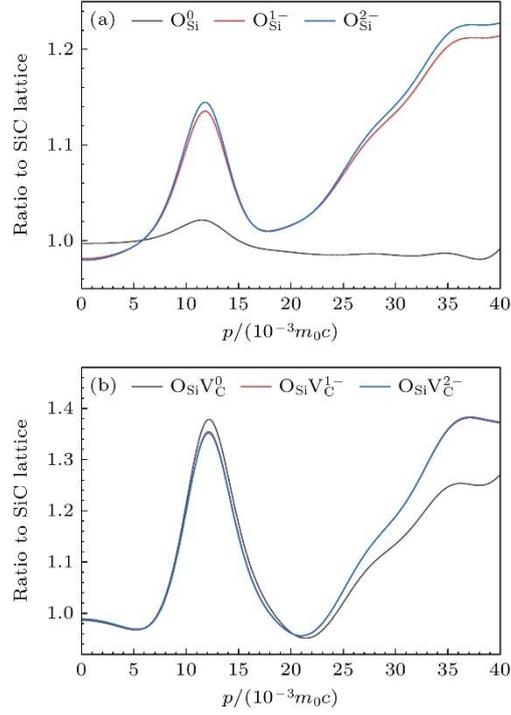

**Figure 5.** Momentum distributions ratio curves of annihilating electron-positron pairs for various charge states of $O_{Si}$ (a) and $O_{Si}V_C$ (b) in 3$C$-SiC.

Fig. 4(a) shows the Doppler spectra of carbon, silicon and oxygen. In Fig. 4(a), the oxygen characteristic peak appears at $p = 13.4 \times 10^{-3} m_0 c$ with a peak value of 3.69, the carbon characteristic peak appears at $p = 12.3 \times 10^{-3} m_0 c$ with a peak value of 2.08, and the silicon characteristic peak appears in $p = 13.8 \times 10^{-3} m_0 c$ with a peak of 1.06. The data of Fig. 4(a) show that carbon, silicon and oxygen show different momentum distribution characteristics, so. In Fig. 4(b), the Doppler spectra of seven 3$C$-SiC defects are shown. The CDB peak positions of $O_{Si}V_C$, $O_C V_{Si}$, $V_{Si}$ and $V_{Si+C}$ are the same, but the peak intensities are significantly different. The CDB peak positions of $V_C$, $O_C$ and $O_{Si}$ are the same. All CDB curves have two peaks at $p = 0$ near $p = 12 \times 10^{-3} m_0 c$, and $V_C$, $V_{Si}$, $V_{Si+C}$ and $O_C V_{Si}$ take a peak with a ratio greater than 1 at $p = 0$, $O_{Si} V_C$, $O_C V_{Si}$ and $O_{Si}$ take a peak with a ratio greater than 1 near $p = 12 \times 10^{-3} m_0 c$, which is consistent with $S$ calculated in Table 2. Where $S_{rel}$ represents the ratio of the $S$ parameter of the defect state to the defect-free state, and $W_{rel}$ represents the ratio of the $W$ parameter of the defect state to the defect-free state. The increase of valence electron annihilation between positron and defect leads to the increase of $S$ parameter of vacancy type defect compared with that of no defect. It is worth noting that the $S_{rel}$ and $W_{rel}$ values of $O_C$ are close to 1.0, indicating that the positron annihilation characteristics of $O_C$ are the same as those of defect-free SiC, which is consistent with the positron annihilation lifetime calculation. In the vacancy-type defects exhibited by Table 2, the $S_{rel}$ parameters of divacancy are larger than the $S_{rel}$ of monovacancy, and the $S_{rel}$ parameters of $V_{Si}$ are larger than the $S_{rel}$ parameters of $V_C$. In the substitutional defects, the $S_{rel}$ parameters and the $W_{rel}$ parameters are both close to 1.0.

**Table 2.** Relative $S_{rel}$ and $W_{rel}$ parameters of intrinsic defects and oxygen-doped defects in 3C-SiC.

| Defect type | $S_{rel}$ | $W_{rel}$ |
|---|---|---|
| $V_C$ | 1.020 | 0.948 |
| $V_{Si}$ | 1.063 | 0.872 |
| $V_{Si+C}$ | 1.082 | 0.790 |
| $O_C$ | 1.000 | 0.999 |
| $O_{Si}$ | 0.997 | 1.009 |
| $O_C V_{Si}$ | 1.025 | 1.002 |
| $O_{Si} V_C$ | 0.988 | 1.228 |

The Doppler spectra of 0, – 1 and – 2 charge states of $O_{Si}$ and $O_{Si}V_C$ were calculated and shown in Fig. 5. The results show that there is no significant difference in the trend of Doppler broadening lines except for $O_{Si}^0$ and $O_{Si}V_C^0$. This indicates that except for the charge state transition from $O_{Si}^0$ to $O_{Si}^{1-}$ and from $O_{Si}V_C^0$ to $O_{Si}V_C^{1-}$, the other charge state changes do not significantly affect the local electronic structure of the defect, which is consistent with the results of Tab. 3 data. The peak value of negative valence state $O_{Si}$ in Fig. 5(a) is larger than that of neutral $O_{Si}$ in the high momentum region, while the peak value of negative valence state $O_{Si}V_C$ in Fig. 5(b) is lower than that of zero valence state $V_C$ in the high momentum region. There is no significant change in the curves of – 2 and – 1 valence states of $O_{Si}$ and $O_{Si}V_C$, which is due to the weakening of positron capture by electrons due to the electrostatic shielding effect of electrons.

**Table 3.** Relative $S_{rel}$ and $W_{rel}$ parameters calculated for various charge states of $O_{Si}$ and $O_{Si}V_C$.

| Defect type | $S_{rel}$ | $W_{rel}$ |
|---|---|---|
| $O_{Si}^0$ | 0.997 | 1.009 |
| $O_{Si}^{1-}$ | 0.984 | 1.087 |
| $O_{Si}^{2-}$ | 0.983 | 1.092 |
| $O_{Si}V_C^0$ | 0.988 | 1.228 |
| $O_{Si}V_C^{1-}$ | 0.990 | 1.210 |
| $O_{Si}V_C^{2-}$ | 0.990 | 1.209 |

## 4. Conclusion

In this work, the intrinsic vacancy defects and oxygen dopant defects in 3C-SiC are studied by density functional theory calculations. The defect formation energies, the electron–positron density distributions, the positron annihilation lifetimes, the CDB spectra and corresponding $S$ and $W$ parameters were calculated to reveal the defect-dependent positron annihilation characteristics. The calculation results show that when the $O_{Si}$ and $O_{Si}V_C$ defects are changed from neutral state to negative valence state, there will be an obvious structural relaxation in the local area, which leads to the local distribution of charge, but when the valence state changes between negative valence states ($-1, -2$), the electronic screening effect of 3$C$-SiC leads to the nonlocal distribution of electron and positron charge density. According to the results of defect formation energy, the positron annihilation lifetime of the stable structures is calculated. It is found that the positron annihilation lifetime calculated by meta-GGA functional, which takes into account the spin polarization effect and orbital kinetic energy density, is 5 — 10 PS larger than that measured in defect-free SiC,. The calculated results of Doppler momentum spectra show that the intrinsic vacancy defects and oxygen doped defects in 3$C$-SiC can be distinguished by Doppler momentum spectra and $S$,$W$ parameters. In $O_{Si}$, the electron screening effect plays a major role, while in $O_{Si}V_C$, the localization of vacancies plays a major role, which leads to the difference of momentum spectrum distribution between different valence states of $O_{Si}$ and $O_{Si}V_C$. The results indicate that the combined use of positron annihilation lifetime and Doppler momentum spectroscopy provides a powerful approach for identifying oxygen-doped defects of SiC in irradiated environments, which has potential value in the detection of radiation damage effects.

**References**

[1] Petti D A, Buongiorno J, Maki J T, Hobbins R R, Miller G K 2003 Nucl. Eng. Des. 222 281

[2] Franceschini F, Ruddy F H 2011 Silicon Carbide Neutron Detectors (Rijeka: InTech) pp275–296

[3] Jiang W L, Jiao L, Wang H Y 2011 J. Am. Ceram. Soc. 94 4127

[4] Fan X J, Ye R Q, Peng Z W, Wang J J, Fan A L, Guo X 2016 Nanotechnology 27 255604

[5] Zhang Y M, Zhu L H, Ban Z G, Liu Y X 2012 Hard Alloy 29 66

[6] Wang P R, Gou Y Z, Wang H 2020 J. Inorg. Mater. 35 525

[7] Ishikawa T, Kohtoku Y, Kumagawa K, Yamamura T, Nagasawa T 1998 Nature 391 773

[8] Ishikawa T 2005 Polymeric and Inorganic Fibers (Berlin, Heidelberg: Springer) (Vol. 178) p109

[9]   Rosso E F, Baierle R J 2013 Chem. Phys. Lett. 568 140

[10] Gali A, Heringer D, Deák P, Hajnal Z, Frauenheim T, Devaty R P, Choyke W J 2002 Phys. Rev. B 66 125208

[11] West R N 1973 Adv. Phys. 22 263

[12] Puska M J, Nieminen R M 1994 Rev. Mod. Phys. 66 841

[13] Zhang L J, Wang L H, Liu J D, Li Q, Cheng B, Zhang J, An R, Zhao M L, Ye B J 2012 Acta Phys. Sin. 61 237805

[14] Zhang H J, Wang D, Chen Z Q, Wang S J, Xu Y M, Luo X H 2008 Acta Phys. Sin. 57 7333

[15] Zhang L J, Zhang C C, Liao W, Liu J D, Gu B C, Yuan X D, Ye B J 2015 Acta Phys. Sin. 64 097802

[16] Hao Y P, Chen X L, Cheng B, Kong W, Xu H X, Du H J, Ye B J 2010 Acta Phys. Sin. 59 2789

[17] Huang S J, Zhang W S, Liu J D, Zhang J, Li J, Ye B J 2014 Acta Phys. Sin. 63 217804

[18] Xu H X, Hao Y P, Han R D, Weng H M, Du H J, Ye B J 2011 Acta Phys. Sin. 60 067803

[19] Liu J D 2010 Ph. D. Dissertation (Hefei: University of Science and Technology of China

[20] Lam C H, Lam T W, Ling C C, Fung S, Beling C D, Hang D S, Weng H M 2004 J. Phys.: Condens. Mat. 16 8409

[21] Staab T E M, Puska M J, Nieminen R M, Torpo L M 2001 Materials Science Forum (Zurich: Trans Tech Publications Ltd) p533

[22] Tuomisto F, Makkonen I 2013 Rev. Mod. Phys. 85 1583

[23] Brauer G, Anwand W, Coleman P G, Knights A P, Plazaola F, Pacaud Y, Skorupa W, Störmer J, Willutzki P 1996 Phys. Rev. B 54 3084

[24] Brauer G, Anwand W, Nicht E M, Kuriplach J, Šob M, Wagner N, Coleman P G, Puska M J, Korhonen T 1996 Phys. Rev. B 54 2512

[25] Kawasuso A, Yoshikawa M, Itoh H, Krause-Rehberg R, Redmann F, Higuchi T, Betsuyaku K 2006 Physica B 376 350

[26] Hu X, Koyanagi T, Katoh Y, Wirth B D 2017 Phys. Rev. B 95 104103

[27] Kresse G, Hafner J 1993 Phys. Rev. B 47 558

[28] Kresse G, Furthmüller J 1996 Phys. Rev. B 54 11169

[29] Kresse G, Furthmüller J 1996 Comput. Mater. Sci. 6 15

[30] Perdew J P, Wang Y 1992 Phys. Rev. B 45 13244

[31] Perdew J P, Kurth S, Zupan A, Blaha P 1999 Phys. Rev. Lett. 82 2544

[32] Verma P, Truhlar D G 2017 J. Phys. Chem. C 121 7144

[33] Hohenberg P, Kohn W 1964 Phys. Rev. 136 B864

[34] Kohn W, Sham L J 1965 Phys. Rev. 140 A1133

[35] Blöchl P E 1994 Phys. Rev. B 50 17953

[36] Rauch T, Munoz F, Marques M A L, Botti S 2021 Phys. Rev. B 104 064105

[37] Zhang H T, Yan L, Tang X, Cheng G D 2024 Phys. Lett. A 525 129888

[38] Levinshtein M E, Rumyantsev S L, Shur M S 2001 Properties of Advanced Semiconductor Materials: GaN, AlN, InN, BN, SiC, SiGe (Hoboken: John Wiley & Sons) pp96–104

[39] Nieminen R M, Boronski E, Lantto L J 1985 Phys. Rev. B 32 1377

[40] Boroński E, Nieminen R M 1986 Phys. Rev. B 34 3820

[41] Arponen J, Pajanne E 1979 Ann. Phys. 121 343

[42] Barbiellini B, Puska M J, Torsti T, Nieminen R M 1995 Phys. Rev. B 51 7341

[43] Kuriplach J, Barbiellini B 2014 J. Phys.: Conf. Ser. 505 012040

[44] Boroński E 2010 Nukleonika 55 9

[45] Asoka-Kumar P, Alatalo M, Ghosh V J, Kruseman A C, Nielsen B, Lynn K G 1996 Phys. Rev. Lett. 77 2097

[46] Alatalo M, Asoka-Kumar P, Ghosh V J, Nielsen B, Lynn K G, Kruseman A C, Van Veen A, Korhonen T, Puska M J 1998 J. Phys. Chem. Solids 59 55

[47] Szpala S, Asoka-Kumar P, Nielsen B, Peng J P, Hayakawa S, Lynn K G, Gossmann H J 1996 Phys. Rev. B 54 4722

[48] Kawasuso A, Maekawa M, Betsuyaku K 2010 J. Phys. Conf. Ser. 225 012027

[49] Kong W, Xi C Y, Ye B J, Weng H M, Zhou X Y, Han R D 2004 High Energy Phys. Nucl. 28 1234

[50] Liu X G, Deng L, Hu Z H, Li R, Fu Y G, Li G, Wang J 2016 Acta Phys. Sin. 65 092501

[51] Alatalo M, Barbiellini B, Hakala M, Kauppinen H, Korhonen T, Puska M J, Saarinen K, Hautojärvi P, Nieminen R M 1996 Phys. Rev. B 54 2397

[52] Makkonen I, Hakala M, Puska M J 2006 Phys. Rev. B 73 035103

[53] Tang Z, Toyama T, Nagai Y, Inoue K, Zhu Z Q, Hasegawa M 2008 J. Phys.: Condens. Matter 20 445203

[54] Wiktor J, Jomard G, Torrent M, Bertolus M 2013 Phys. Rev. B 87 235207

[55] Kawasuso A, Itoh H, Morishita N, Yoshikawa M, Ohshima T, Nashiyama I, Okada S, Okumura H, Yoshida S 1998 Appl. Phys. A 67 209

[56] Panda B K, Brauer G, Skorupa W, Kuriplach J 2000 Phys. Rev. B 61 15848

[57] Kawasuso A, Maekawa M, Fukaya Y, Yabuuchi A, Mochizuki I 2011 Phys. Rev. B 83 100406